\documentclass[fleqn,usenatbib]{mnras}

\usepackage{graphicx}	
\usepackage{amsmath}	
\usepackage{amssymb}	
\usepackage[dvipsnames]{xcolor}
\usepackage{hyperref}

\usepackage{newtxtext,newtxmath}
\usepackage{ulem}
\usepackage[T1]{fontenc}

\DeclareRobustCommand{\VAN}[3]{#2}
\let\VANthebibliography\thebibliography
\def\thebibliography{\DeclareRobustCommand{\VAN}[3]{##3}\VANthebibliography}

\newcommand{\dif}{\mathrm{d}}

\begin{document}



\title[Fallback discs around millisecond magnetars]{Evolution of fallback discs around millisecond magnetars: effect of supercritical accretion on GRB afterglows}

\author[S. \c{C}\i{}k\i{}nto\u{g}lu et al.]{
Sercan \c{C}\i{}k\i{}nto\u{g}lu$^{1}$\thanks{E-mail: cikintoglus@itu.edu.tr},
Sinem {\c S}a{\c s}maz$^{1}$,
M. Hakan Erkut$^{2}$,
K. Yavuz Ek\c{s}i$^{1}$
\\
$^{1}$Istanbul Technical University, Faculty of Science and Letters,
Physics Engineering Department, 34469, İstanbul, TURKEY  \\
$^{2}$Boğaziçi University, Department of Physics, 34342 Bebek, İstanbul, TURKEY}

\date{Accepted XXX. Received YYY; in original form ZZZ}

\pubyear{2022}

\label{firstpage}
\pagerange{\pageref{firstpage}--\pageref{lastpage}}
\maketitle

\begin{abstract}
Some models of gamma-ray bursts (GRBs) invoke nascent millisecond magnetars as the central engine and address the X-ray afterglows with the interaction of magnetar magnetospheres with fallback discs. We study the evolution of fallback discs interacting with the millisecond magnetars. Initially, the accretion rate in the fallback disc is very high, well above the rate required for the Eddington limit. The inner parts of such a disc, even if it is cooling by the neutrino emission, get spherical due to the radiation pressure, which regulates the mass accretion rate within the spherization radius. Such a disc can not penetrate the light cylinder radius for the typical magnetic fields, and the initial spin frequencies invoked for the magnetars. As a result of the auto-regulation of the accretion flow, the fallback disc can not interact directly with the magnetar's magnetosphere within the first few days. This has implications for the fallback disc models of GRB afterglows since the accretion and propeller luminosities, in the presence of radiation pressure, are too low to address the typical luminosities of X-ray afterglows. 
\end{abstract}

\begin{keywords}
accretion, accretion discs -- stars: neutron -- gamma-ray burst: general
\end{keywords}



\section{Introduction}

Gamma-ray bursts (GRBs) occurring at cosmological distances are the most energetic events in the Universe \citep[see, e.g.,][]{Ghi09, Ber14, Dav15, Kumar15}. Nascent neutron stars with millisecond rotation periods and very strong magnetic fields, i.e., millisecond magnetars, are likely involved in GRBs \citep{dun+92,Usov92,Dai98a,zha+01,met+11}. 
Several models invoke millisecond magnetars as the central engine or contributing energy source of GRBs via interaction of the star's highly collimated relativistic wind with the circumburst environment \citep{mes+97,sar+98}, extraction of the rotational energy \citep{Usov92,Dai98a,zha+01}, or accretion from a disc that is formed around the star \citep{zha+10, Dai12, Metzger18}.

GRBs are initially observed in the gamma-ray band, so-called prompt emission. A peak luminosity in the range $10^{47}-10^{54}\,\mathrm{erg\,s^{-1}}$ is inferred for GRBs with known redshifts \citep[see, e.g.,][]{Wang20}. 
The prompt emission is followed by the afterglow emission in lower energy bands \citep{cos+97,van+97,fra+97}. The afterglows of several GRBs have been observed from seconds to thousands of seconds \citep[see, e.g.,][]{eva+09,lia+12,cha+12}.
In the standard model of GRBs, the afterglow emission and its evolution are mainly due to the interaction of the decelerating relativistic ejecta (usually called a fireball) with the surrounding medium, which causes external shocks that dissipate energy \citep{mes+97, sar+98}. The light curves of the X-ray afterglows are composed of power-law components with different temporal decay indices, usually a steep decay followed by a shallow decay which is subsequently followed by another decay phase \citep[see, e.g.,][]{nou+06,pan06,zha+06}. These components are explained as the result of the delayed radiation from higher latitudes due to the curvature effect after the decline of the source of the prompt emission \citep{fen+96,kum+00}, energy injection into the forward shock \citep{ree+98,sar+00,ree+00,nou+06,pan06,met+11}, subsequent decay of the energy injection, and evolution, orientation or geometry of the relativistic beaming \citep{mes+98,rho+99,lip+01}. The shallow decay phase, often called the plateau phase, has drawn interest as it might suggest additional mechanisms, and therefore it may require an extension of the fireball model.
Some explanations for the origin of the plateau phase invoke time evolution of shock parameters such as electron energy fraction and magnetic energy fraction \citep{iok+06}, emission originating from a reverse shock \citep{gen+07, Uhm07}, 
emission from a jet that has an angular structure (i.e., a structured jets) including effects contributed by the angle between the line of sight and jet components with possible contribution from the tail of the prompt emission \citep{tom+06,eic+06},
``late-prompt'' emission due to prolonged central engine activity \citep{ghi+07}, external shock emission involving structured jets combined with off-axis view \citep{ben+20}, high-latitude prompt emission from a structured jet \citep{oga+20},
the rotational energy of a millisecond magnetar 
\citep[see, e.g.,][]{Dai98a,Dai98b,zha+01,fan06,row+10,dal+11,Lasky17,Sasmaz19,Cikintoglu20,Suvorov20},
and fallback accretion onto a black hole \citep{kum+08,lin+10,Cannizzo11}. \citet{Dereli22} has recently argued that the fireball model can explain the plateau phase with the environmental and dynamical conditions specific to GRB progenitors, namely high-mass stars, without additional mechanisms. 

One of the models invoked for addressing the GRB afterglows invokes the interaction of a fallback disc with a millisecond magnetar \citep{Dai12}. This model can fit the GRB afterglow light curves remarkably well if high mass accretion onto the star well above the Eddington limit is invoked
\citep{Gompertz14,Yu15,Gibson17,Gibson18,Lin20}. 
The purpose of the present paper is to examine the evolution of such fallback discs, emphasising the radiation pressure's role in regulating the mass inflow rate at the inner disc and the luminosity that can be attained.

The magnetosphere of a neutron star interacts with the disc in three different regimes depending on the location of the inner radius of the disc relative to the corotation and the light cylinder radii \citep{Lipunov92}. If the mass flow rate in the disc, $\dot{M}$, is sufficiently high, the flow overcomes the magnetic barrier. It gets inside the corotation radius, defined as the radial location at which the angular velocity of the disc equals the star's angular frequency $\Omega_*$ (see equation~\ref{corotation}), and can accrete onto the star (accretion phase). If the stellar magnetic field is sufficiently strong, it stops the disc beyond the light cylinder, defined as the radius at which the rigidly rotating closed stellar magnetic field lines would reach the speed of light (see equation~\ref{light_cylinder}). In this case, the disc can not apply torque on the star, and the star spins down under the magnetic dipole radiation torque alone, corresponding to the so-called ejector phase.
If the inner radius of the disc is between the corotation radius and the light cylinder, the system is in the propeller phase. In this case, the magnetic field lines expel the flow, and the star spins down \citep{Shvartsman70, Illarionov75}.

Initially, the accretion rate in the disc is expected to be very high, and the disc is possibly cooling by neutrino emission \citep{Popham99,Narayan01}. According to the 1D simulations of \citet{Chen07}, the neutrino cooling becomes inefficient when the accretion rate drops below $\sim 10^{-3}\, M_{\sun}/{\rm s}$. 
The system would be very bright due to accretion onto the star.
Moreover, at such high mass inflow rates, the spin-down power transferred to the disc can make it very bright, even if the system is in the propeller stage \citep{Eksi05,Erkut19}.
In some works, it has been argued that this exposes very high luminosities sufficient to power supernovae (SN) \citep{Piro11,Lin20,Lin21} or GRBs \citep{Dai12,Gompertz14,Yu15,Gibson17,Gibson18,Metzger18,Lin20,Li21,Meredith23}.

The disc is not expected to be geometrically thin for high mass-accretion rates exceeding the Eddington limit. The radiation pressure in the disc is substantial, even if its cooling is dominated by neutrino emission. 
Thus the inner part of such a disc gets thicker, and the inflow is regulated such that the mass accretion rate is a few times the Eddington rate \citep{Shakura73,Lipunov82,Lipunova99}. 
The spherization of the inner disc by radiation pressure was considered for the early evolution of fallback discs in the context of fallback discs around anomalous X-ray pulsars \citep{Eksi03}. 
As a result of this regulation, the inner radius of the disc becomes independent of the mass-accretion rate.
Therefore, the flow may not be able to overcome the magnetic barrier of the stellar field despite its high mass-accretion rate.
Moreover, the produced luminosity can achieve only a few factors of the Eddington limit thanks to the regulation of the mass inflow rate.

Recently, a detailed model was proposed for the supercritical regime of disc-magnetosphere interaction in the context of ultra-luminous X-ray sources in X-ray binaries \citep{Erkut19}.
In the present paper, we investigate the implications of this model for fallback discs around nascent neutron stars and fallback disc models of GRB afterglows. We show that if the mass inflow rate in the disc is regulated in the super-critical regime as conceived by \citet{Shakura73}, achieving the arbitrarily high luminosities required to power SN and GRB afterglows is impossible.

In Sec.~\ref{sec:supercritical}, we introduce the model for the super-critical regime of fallback discs. We then describe the evolution of the interaction of the disc with a millisecond magnetar in Sec.~\ref{sec:fallbackdisk}. 
Finally, we estimate the possible luminosity of the system in different modes,
and discuss its possible implications in the context of X-ray afterglows in Sec.~\ref{sec:luminosity}.

\section{Super-critical Regime}
\label{sec:supercritical}
The torque acting on the star due to the disc-magnetosphere interaction is
\begin{equation}
N_{\rm disc} = n\dot{M}_{\rm in} \sqrt{G M R_{\rm in}},
\end{equation}
where $M$, $\dot{M}_{\rm in}$, $R_{\rm in}$, and $n$ are the mass of the neutron star, the mass-accretion rate in the innermost disc, the inner radius of the disc, and the dimensionless torque, respectively.
The dimensionless torque $n$ depends on the fastness parameter $\omega_*$, which is defined as the ratio of the star's spin to
the Keplerian angular velocity at the inner disc radius, $\omega_*\equiv \Omega_*/\Omega_{\rm K}(R_{\rm in})$ with $\Omega_{\rm K}=\sqrt{GM/r^3}$.
To account for a thick disc deviating from the Keplerian rotation, we define $\omega_{\rm p}\equiv \Omega(R_{\rm in})/\Omega_{\rm K}(R_{\rm in})$. The dimensionless torque can be expressed as
\begin{equation}
n\left(\omega_*\right)=n_0\left(1-\frac{\omega_*}{\omega_p}\right)\,,
\end{equation}
with $n_0$ being a dimensionless constant of order unity \citep{Erkut19}.
On the other hand, the star also experiences a spin-down torque due to the magnetic dipole radiation but slightly modified by the presence of the disc \citep{Parfrey16,Parfrey17},
\begin{equation}
N_{\rm dip}=
\begin{cases}
\displaystyle
-\frac{\mu^2\Omega_*^3}{c^3} \left(\frac{R_{\rm LC}}{R_{\rm in}}\right)^2,& \qquad R_{\rm in}<R_{\rm LC}\,,\\
\displaystyle
-\frac{\mu^2\Omega_*^3}{c^3} ,& \qquad R_{\rm in}>R_{\rm LC}\,.
\end{cases}
\label{eq:Ndip}
\end{equation}
Here, $\mu$ is the magnetic dipole moment of the neutron star, $R_{\rm LC}$ is the light-cylinder radius, and $c$ is the speed of light.
Note that, for simplicity, we assume the star's magnetic field is aligned with the star's rotational axis. 
The spin-down luminosity due to the loss of the rotational kinetic energy is given by
\begin{equation}
L_{\rm sd} = -I\Omega_*\dot{\Omega}_* = -\Omega_*\left(N_{\rm dip}+N_{\rm disc}\right)\,,
\label{eq:lumsd}
\end{equation}
which vanishes when the star is spinning up.

The mass accretion rate within the fallback disc is given by
\begin{equation}
\dot{M}_{0}=\frac{2M_{\rm fb}}{3t_{\rm fb}}\left(1+\frac{t}{t_{\rm fb}}\right)^{-5/3},
\end{equation}
where $M_{\rm fb}$ is the mass of the fallback disc and $t_{\rm fb}$ is the characteristic fallback time \citep{Michel88,Metzger18}.
The super-critical accretion rate of the flow, $\dot{M}_0>M_{\rm Edd}\equiv L_{\rm Edd}/(\epsilon c^2)$, 
causes the spherization of the disc inside a critical radius, so-called the spherization radius,
\begin{equation}
R_{\rm sp}=\frac{27\epsilon\sigma_{\rm T}\dot{M}_0}{8\pi m_{\rm p} c}
\simeq 1.43\times 10^{9}\;\epsilon\left(\frac{\dot{M}_0}{10^{20}\,\mathrm{g}\,\mathrm{s^{-1}}}\right)\,{\rm cm}, \label{spr}
\end{equation}
where $\sigma_{\rm T}$ is the Thomson cross-section and
$m_{\rm p}$ is the mass of a proton \citep{Shakura73}. Here,
\begin{equation}
\epsilon= \frac{GM}{Rc^2} \simeq 0.2\,\,M_{1.4}R_{10}^{-1} 
\label{eq:efficiency}
\end{equation}
is the efficiency coefficient for a neutron star of mass $M$ and radius $R$ where we used the notation $M_{1.4}=M/(1.4M_{\sun})$ and $R_{10}=R/(10\,\mathrm{km})$. Within the spherization radius, the inflow is regulated as
\begin{equation}
\dot{M}=
\begin{cases}
\displaystyle
\dot{M}_0 \frac{r}{R_{\rm sp}}\,, & \qquad r<R_{\rm sp}\,, \\
\dot{M}_0\,, & \qquad r>R_{\rm sp}\,. 
\end{cases}
\end{equation}
Note that $\dot{M}_{\rm in}$ is given by $\dot{M}$ at $r=R_{\rm in}$. If $R_{\rm in}<R_{\rm sp}$, $\dot{M}_{\rm in}=(R_{\rm in}/R_{\rm sp})\dot{M}_0$ and if $R_{\rm in}>R_{\rm sp}$, $\dot{M}_{\rm in}=\dot{M}_0$. While the inner radius of the disc is conventionally proportional to the Alfv\'{e}n radius beyond the spherization radius, it is substantially different inside the spherization radius, as suggested by
\begin{equation}
R_{\rm in}=
\begin{cases}
\displaystyle
\left(\frac{\gamma_{\rm p} \omega_{\rm p} \mu^2 R_{\rm sp}\delta}{n_0\dot{M}_0\sqrt{GM}}\right)^{2/9}, &
\qquad  \hspace{-10pt}R_{\rm co}<R_{\rm in}<R_{\rm sp}\,, \\
\displaystyle
\left(\frac{\mu^2 R_{\rm sp}\delta}{\dot{M}_0\sqrt{GM}}\right)^{2/9}, &
\qquad  \hspace{-10pt}R_{\rm in}<R_{\rm co}\quad\text{and}\quad R_{\rm in}<R_{\rm sp}\,, \\
\displaystyle
\left(\frac{\mu^2 \delta}{\dot{M}_0\sqrt{GM}}\right)^{2/7},&
\qquad  \hspace{-10pt}R_{\rm sp}<R_{\rm in}\,, \label{rin}
\end{cases}
\end{equation}
where $R_{\rm co}$ is the corotation radius, $\delta$ is the relative width of the coupled domain between the disc and the magnetosphere and $\gamma_{\rm p}$ is the constant coefficient of the azimuthal pitch in the propeller phase \citep{Erkut19, Erkut20}. The first and second cases in equation~(\ref{rin}) correspond to the supercritical propeller and accretion regimes, respectively. In both supercritical regimes with $R_{\rm in}<R_{\rm sp}$, the dependence of the inner disc radius on the stellar magnetic dipole moment is $R_{\rm in}\propto \mu^{4/9}$ whereas the dependence switches to $R_{\rm in}\propto \mu^{4/7}$ in the third case of equation~(\ref{rin}). The latter corresponds to the subcritical accretion regime with $R_{\rm in}$ exceeding the spherization radius but still staying inside the corotation radius. In all these three cases, the source is expected to be luminous. The source becomes faint only when $R_{\rm in}$ exceeds both $R_{\rm sp}$ and $R_{\rm co}$. The two expressions for the inner disc radius in the supercritical propeller and accretion regimes in equation~(\ref{rin}) can be seen to become identical if $\gamma_{\rm p}\omega_{\rm p}/n_0$ is chosen to be 1. Such a choice reduces the number of unknown constant parameters and provides a natural and smooth transition from the propeller to the accretion phase.

The potential energy released by the infalling matter on the surface of the neutron star results in an accretion luminosity of
\begin{equation}
L_{\rm acc}=\frac{GM\dot{M}_{\rm in}}{R}-\frac{GM\dot{M}_{\rm in}}{R_{\rm in}},
\end{equation}
which otherwise vanishes in the propeller regime as the accretion halts.
Additionally, the gravitational potential energy is converted to luminosity as a result of the dissipative processes within the disc,
\begin{equation}
L_{\rm grav}=\int^{\infty}_{R_{\rm in}} \frac{GM\dot{M}}{r^2}\dif r
=
\begin{cases}
\displaystyle
\frac{2L_E}{27\epsilon}\left[
\ln\left(\frac{R_{\rm sp}}{R_{\rm in}}\right)+1\right],
&\quad R_{\rm in}<R_{\rm sp}\,, \\
\displaystyle
\frac{GM\dot{M}_0}{R_{\rm in}}=
\frac{2L_{\rm Edd}}{27\epsilon}\frac{R_{\rm sp}}{R_{\rm in}},
&\quad R_{\rm in}>R_{\rm sp}\,.
\end{cases}
\end{equation}
On the other hand, due to outflowing matter in the inner region, the energy is lost at the rate,
\begin{equation}
L_{\rm out}=\frac{1}{2}\int v^2_{\rm out}\,\dif\dot{M}
=
\begin{cases}
\displaystyle
-\frac{2L_{\rm Edd}}{27\epsilon}\left(\frac{L_{\rm tot}}{L_{\rm Edd}}-1\right)\ln\frac{R_{\rm sp}}{R_{\rm in}},&
\quad R_{\rm in}<R_{\rm sp}\,, \\
\displaystyle
-\frac{1}{2}\dot{M}_0 v^2_{\rm out}\,,
&\quad R_{\rm in}>R_{\rm sp}\,,
\end{cases}
\end{equation}
where the velocity of the outflow is $v_{\rm out}=R_{\rm in}\Omega_{\rm K}\left(1-n\right)$ \citep{Eksi05,Erkut19}. 
The outflow luminosity vanishes if the outflow velocity is smaller than the escape velocity, i.e., $v_{\rm out}<v_{\rm esc}\equiv \sqrt{2GM/R}$.
The total luminosity is given by the sum of the luminosities of these four different mechanisms,
\begin{equation}
L_{\rm tot}=L_{\rm acc}+L_{\rm sd}+L_{\rm grav}+L_{\rm out}.
\end{equation}

\begin{figure*}
\includegraphics[width=0.9\textwidth]{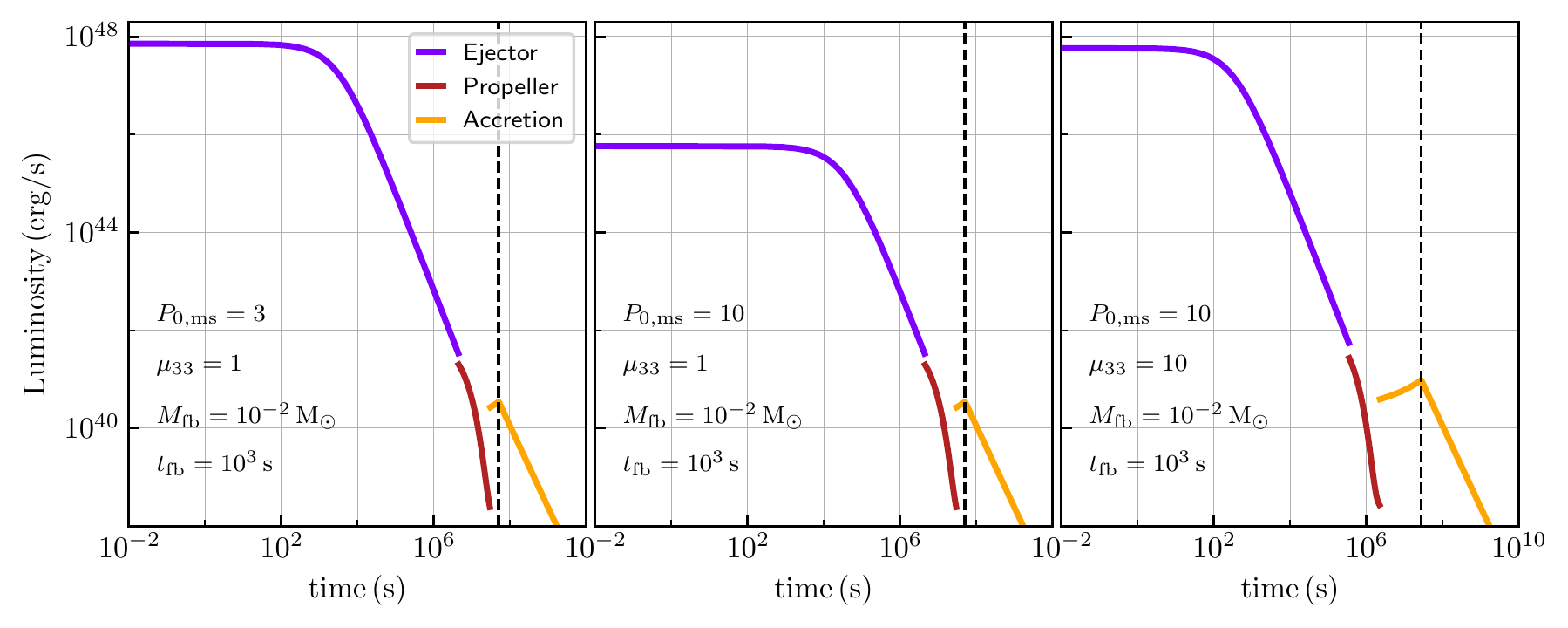}
\caption{The luminosity produced for three different parameter sets. We set $M_{1.4}=R_{10}=\delta_{0.01}=1$, and $b=1$ for all three cases. The black dashed lines mark the time when the system switches to subcritical accretion.
}
\label{fig:lum}
\end{figure*}

\section{Evolution of the fallback disc}
\label{sec:fallbackdisk}
In the supercritical regime, the inner disc radius can be estimated in line with equation~(\ref{rin}) as
\begin{equation}
R_{\rm in}\simeq 5.5\times 10^8\,\mu_{33}^{4/9} M_{1.4}^{1/9} R_{10}^{-2/9} \delta_{0.01}^{2/9}\,{\rm cm}. \label{rinum}
\end{equation}
where $\mu_{33}=\mu/(10^{33}\,\mathrm{G\,cm^3})$, and $\delta_{0.01}=\delta/0.01$.
Similarly, the characteristic radii (the corotation radius and the light cylinder radius) can be expressed as
\begin{equation}
R_{\rm co} = \left(\frac{GM}{\Omega_*^2}\right)^{1/3}= 1.7\times 10^6\,
M_{1.4}^{1/3} P_{\rm ms}^{2/3}\,\mathrm{cm}\,, \label{corotation}
\end{equation}
and
\begin{equation}
R_{\rm LC} =\frac{c}{\Omega_*}=4.8\times 10^6\,P_{\rm ms}\,\mathrm{cm}\,, \label{light_cylinder}
\end{equation}
where $P_{\rm ms}=P/(1\,\mathrm{ms})$.
Accordingly, the critical periods at which the inner radius of the disc would coincide with the corotation and the light cylinder radii are, respectively,
\begin{align}
P_{\rm co, ms} =&\, 5819\, \mu_{33}^{2/3} M_{1.4}^{-1/3}R_{10}^{-1/3}\delta_{0.01}^{1/3}\,, \\
P_{\rm LC, ms} =&\, 115\, \mu_{33}^{4/9}M_{1.4}^{1/9}R_{10}^{-2/9}\delta_{0.01}^{2/9}\,.
\end{align}
Therefore, the fallback disc initially has to be beyond light-cylinder radius 
if the new born object after a core-collapse supernova is a millisecond magnetar, i.e., $\mu>10^{33}\,\mathrm{G\,cm^3}$ and $P<100\,\mathrm{ms}$.

Until the magnetar slows down sufficient to the the inner radius of the
disc gets inside the corotation radius, the star slows down
under solely magnetic dipole spin-down torque,
\begin{equation}
\frac{\dif\Omega_*}{\dif t}=-\frac{\mu^2\Omega_*^3}{Ic^3}\,,
\end{equation}
and its spin evolves as
\begin{equation}
\Omega_* = \Omega_0 \left(1+\frac{2\mu^2\Omega_0^2}{Ic^3}t\right)^{-1/2}\,,
\end{equation}
which implies
\begin{equation}
P=P_0\left(1+\frac{8\pi^2\mu^2}{P_0^2 Ic^3}t\right)^{1/2}\,.
\end{equation}
We can estimate the required time for commencing of the propeller regime.
For instance, it is at the order of $10^6\,\mathrm{s}$ 
if the dipole magnetic moment of the magnetar is $\mu_{33}=1$ and 
its period is in the range of $1\,\mathrm{ms}<P<10\,\mathrm{ms}$.

After the system switches to the propeller regime, the star slows down much rapidly 
as its interaction with the disc exerts additional negative torque. 
Finally, the system would switch to the accretion regime 
when the corotation radius moves beyond than the inner radius of the disc.

\section{The luminosity of the system and its implication for the X-ray light curves of the GRB afterglows}
\label{sec:luminosity}
When the disc is beyond the light cylinder, only the dipole radiation powers the luminosity of the system
and it can be written as
\begin{equation}
L_{\rm dipole}= 5\times 10^{49}\,\mu_{33}^2 P_{\rm ms}^{-4}\,\mathrm{erg\,s^{-1}}\,,
\end{equation}
by using equations \eqref{eq:Ndip} and \eqref{eq:lumsd}.
Accordingly, we can deduce the total luminosity of the system at the commencement of the propeller regime as
\begin{align}
    L_{\rm prop, max}=&\, 5\times 10^{49}\,\mu_{33}^2 P_{\rm LC, ms}^{-4}\,\mathrm{erg\,s^{-1}} 
    \notag \\
    =&\,3\times 10^{42}\, \mu_{33}^{2/9}M_{1.4}^{-4/9}R_{10}^{8/9}\delta_{0.01}^{-8/9}\,\mathrm{erg\,s^{-1}}\,.
\end{align}
Note that, since the luminosity decreases by the time in the propeller regime (see Fig.~\ref{fig:lum}), this also corresponds the maximum luminosity
which can be generated in the propeller regime. Therefore, 
the luminosity of the system can achieve mostly $\sim 10^{42}-10^{43}\,\mathrm{erg\,s^{-1}}$ 
considering magnetic field of a magnetar, i.e., $\mu_{33}\sim 1-10^3$.

On the other hand, the dominant component of the X-ray luminosity of an accreting system is due to accretion. 
Accordingly, we estimate the total luminosity as
\begin{equation}
L_{\rm tot}\simeq L_{\rm acc}\simeq\frac{GM\dot{M}_{\rm in}}{R}
=\frac{8\pi m_{\rm p}c^3}{27\sigma_{\rm T}}R_{\rm in} 
\end{equation}
in the supercritical accretion regime by using $\dot{M}_{\rm in}=(R_{\rm in}/R_{\rm sp})\dot{M}_0$ and $\epsilon=GM/(Rc^2)$ with equation~(\ref{spr})
and we find
\begin{equation}
L_{\rm tot}\simeq 6.3\times 10^{39}\,\frac{R_{\rm in}}{10^8\,{\rm cm}}\,{\rm erg\,s^{-1}}\,. \label{ltot}
\end{equation}
Since the efficiency coefficient $\epsilon \simeq 0.2$ for a neutron star (see equation~\ref{eq:efficiency}), the supercritical regime with $R_{\rm in}<R_{\rm sp}$ is satisfied if the mass accretion rate $\dot{M}_0$ within the fallback disc sufficiently exceeds $10^{20}$\,g\,s$^{-1}$ (equation~\ref{spr}). The luminosity of the system is at most $\sim 10^{42}\,\mathrm{erg\,s^{-1}}$ (see Fig.~\ref{fig:lum}) 
even if we set the magnetic dipole moment to $\mu_{33}=10^3$, the maximum value a neutron star can hold, while other parameters being fixed at typical values, i.e., $M_{1.4}=R_{10}=\delta_{0.01}=1$ (equations~\ref{rinum} and \ref{ltot}). If, on the other hand, the system is in the subcritical accretion regime with $R_{\rm in}>R_{\rm sp}$, we estimate the total luminosity of the system as
\begin{equation}
L_{\rm tot}\simeq L_{\rm acc}\simeq\frac{GM\dot{M}_{\rm in}}{R}
=\frac{GM\dot{M}_0}{R}
\end{equation}
using $\dot{M}_{\rm in}=\dot{M}_0$. We obtain
\begin{equation}
L_{\rm tot}\simeq 1.9\times 10^{40}\,\frac{M_{1.4}}{R_{10}}\left(\frac{\dot{M}_0}{10^{20}\,\mathrm{g}\,\mathrm{s^{-1}}}\right)\,{\rm erg\,s^{-1}}, \label{totl}
\end{equation}
which can attain values as high as the typical luminosity values of the GRB X-ray afterglows only for unrealistically large mass inflow rates such as $\dot{M}_0\gtrsim 10^{28}$\,g\,s$^{-1}$ in the disc. Even if such extremely high values of $\dot{M}_0$ are possible, they cannot be used in equation~(\ref{totl}) as the subcritical accretion phase is realizable solely for
\begin{equation}
\dot{M}_0\lesssim 1.9\times 10^{20}\,\mu_{33}^{4/9} \delta_{0.01}^{2/9} M_{1.4}^{-8/9} R_{10}^{7/9}\,{\rm g\,s^{-1}}
\end{equation}
resulting from the condition of $R_{\rm in}>R_{\rm sp}$ being satisfied in line with equations~(\ref{spr}) and (\ref{rin}).

Besides the fact that the disc initially would be beyond the light cylinder radius with such strong magnetic fields of millisecond magnetars,
these extreme condition is still insufficient to produce the typical luminosities of the X-ray afterglows following a GRB,
which are of the order of $10^{48}\,\mathrm{erg\,s^{-1}}$, neither in the propeller nor the accretion regimes.
Therefore, the interaction of the star with the fallback disc can be relevant for the X-ray light curves only in the late time
when the luminosity drops a few orders.
On the other hand, if the initial luminosity of the system is $10^{48}\,\mathrm{erg\,s^{-1}}$ 
which can be easily produced by the magnetic dipole radiation of a millisecond magnetar,
such a drop in the luminosity would correspond to 6 orders of magnitude drop in the X-ray flux, which is wider than the sensitivity limit of Swift's XRT\footnote{See \url{https://swift.gsfc.nasa.gov/about\_swift/xrt\_desc.html}} which works between $2\times10^{-14}-9\times10^{-10}\,\mathrm{erg\,cm^{-2}\,s^{-1}}$. Given that the largest luminosity drop detectable by Swift's XRT is $\sim 10^5$ fold, it is implausible to be able to observe a system switching to the propeller or accretion phases for the usual X-ray afterglows that initially have a
luminosity of the order of $10^{48}\,\mathrm{erg\,s^{-1}}$. 

Typically, the observations of the X-ray afterglows are limited to $10^6\,\mathrm{s}-10^7\,\mathrm{s}$.
This time limit and the luminosity constraint require a very strong magnetic dipole moment even for a magnetar. 
For instance, the magnetic dipole moment of the star should be greater than $10^{33}\,\mathrm{G}$
and the initial period should be smaller than $100\,\mathrm{ms}$ for switching to the propeller regime before $10^6\,\mathrm{s}$ as well as
the initial luminosity to be at the order of $10^{48}\,\mathrm{erg\,s^{-1}}$.

The beaming factor amplifies the equivalent luminosity,
\begin{equation}
L_{\rm obs}=L_{\rm tot}/b\,,
\end{equation}
\citep{King09,Bucciantini12}. So, a higher luminosity can be observed with the same set of parameters as the beaming factor decreases despite the fallback disc producing the same power output. Also, the star can initially have a smaller spin frequency since the same amount of luminosity can be observed from the magnetic dipole radiation with a smaller beaming factor. Thus, small values of the beaming factor can solve the detection problem described in the previous paragraphs.
Still, an observed luminosity of $\sim 10^{48}\,\mathrm{erg\,s^{-1}}$ cannot be produced by the disc-magnetosphere interaction,
unless the beaming factor is extreme, i.e.\ as small as $10^{-6}$.
Therefore, the contribution from the fallback disc is not powerful enough to be effective in the early stages of the X-ray afterglows. On the other hand, the propeller phase might be effective during the steep decay of the X-ray light curve following the plateau phase. Moreover, the accretion from a fallback disc may exhibit a second plateau phase with a low luminosity in the late stages of the X-ray light curves. 

\section{Conclusion}

We studied the modes of interaction of a fallback disc with the magnetosphere of a millisecond magnetar, considering the radiation pressure within the disc. Fallback models were recently invoked to address the X-ray afterglows of GRBs, but without considering the regulation of the mass accretion rate due to radiation pressure.

We showed that in the early stage of the system, the fallback disc should be beyond the light cylinder 
due to the strong magnetic field and the short spin period of the millisecond magnetar
and cannot interact with the star until the star slows down sufficiently.
Once the star is slow enough that the disc can intrude into the light cylinder,
the propeller regime commences. Since the propeller torque is stronger than the magnetic dipole torque, the star slows down more rapidly in this stage, and the system soon switches to the accretion regime.

The maximum luminosity that can be generated in the propeller and the accretion regimes
are at least six orders lower than the typical luminosity of X-ray afterglows of GRBs ($\sim 10^{48}\,\mathrm{erg~s^{-1}}$). Since Swift's XRT is sensitive to a flux range of 5 orders of magnitude, the X-ray flux 
would remain below the detection limit after these regimes commence. If the beaming factor is considered, the luminosity can be high enough to be observed. However, it would still remain below the typical luminosity of X-ray afterglows of GRBs unless the beaming factor is extremely small, i.e.,~$b<10^{-6}$. 
The contribution of the disc's presence to the X-ray light curve would be
significant only in the late time and at low luminosity values compared to the early phase of the X-ray light curve.
Concluding these facts and considering the observations, 
the X-ray afterglows of long GRBs are unlikely to be powered by accretion onto a millisecond magnetar from a fallback disc.

\section*{Acknowledgements}
SS and KYE acknowledge support from the Scientific and Technological Research Council of Turkey (TÜBİTAK) with project number 118F028.

\section*{Data availability}

This is a theoretical paper that does not involve any new data. The
model data presented in this article are all reproducible.


\bibliographystyle{mnras}
\bibliography{refs.bib} 



\bsp	
\label{lastpage}
\end{document}